\documentclass[a4paper]{jpconf}

\usepackage{graphicx}
\usepackage{aeguill}
\usepackage{amsmath}
\usepackage{verbatim}

\begin{document}
\title{Differential energy measurement between He- and Li-like uranium intra-shell transitions}

\author{M.~Trassinelli$^1$, A.~Kumar$^2$, H.F.~Beyer$^3$, P.~Indelicato$^4$, R.~M\"artin$^{3,5}$, R.~Reuschl$^1$, Y.S.~Kozhedub$^6$, C.~Brandau$^3$, H.~Br\"auning$^3$,  S.~Geyer$^3$, A.~Gumberidze$^7$, S.~Hess$^3$, P.~Jagodzinski$^8$, C.~Kozhuharov$^3$, D.~Liesen$^3$,  U.~Spillmann$^3$, S.~Trotsenko$^{9}$, G.~Weber$^{3,5}$, D.F.A.~Winters$^{3,5}$, Th.~St\"ohlker$^{3,5,9}$}

\address{$^1$ Institut des NanoSciences de Paris; CNRS; Univ. Pierre et Marie Curie - Paris 6, Paris, France} 
\address{$^2$ Nuclear Physics Division, Bhabha Atomic Research Centre, Mumbai, India} 
\address{$^3$ GSI Helmholtzzentrum f\"ur Schwerionenforschung GmbH, Darmstadt, Germany}
\address{$^4$ Laboratoire Kastler Brossel, \'Ecole Normale Sup\'erieure; CNRS; Univ. Pierre et Marie Curie - Paris 6, Paris, France}
\address{$^5$ Physikalisches Institut, Univ. Heidelberg, Heidelberg, Germany}
\address{$^6$ Department of Physics, St. Petersburg State Univ., St.~Petersburg, Russia}
\address{$^7$ ExtreMe Matter Institue, Darmstadt, Germany}
\address{$^8$ Institute of Physics, Jan Kochanowski Univ., Kielce, Poland}
\address{$^{9}$ Helmholtz-Institut Jena, Jena, Germany}

\ead{martino.trassinelli@insp.jussieu.fr}

\begin{abstract}
We present the first clear identification and highly accurate measurement of the intra-shell transition $1s2p\, ^3\!P_2 \to 1s2s\, ^3\!S_1$ of He-like uranium performed via X-ray spectroscopy.
The present experiment has been conducted at the gas-jet target of the ESR storage ring in GSI (Darmstadt, Germany) where a Bragg spectrometer, with a bent germanium crystal, and a Ge(i) detector were mounted.  
Using the ESR deceleration capabilities, we performed a differential measurement between the $1s2p\, ^3\!P_2 \to 1s2s\, ^3\!S_1$ He-like U transition energy, at 4510 eV, and the $1s^22p\ ^2\!P_{3/2 } \to 1s^22s\, ^2\!S_{1/2}$ Li-like U transition energy, at 4460 eV. 
By a proper choice of the ion velocities, the X-ray energies from the He- and Li-like ions could be measured, in the laboratory frame, at the same photon energy.
This allowed for a drastic reduction of the experimental systematic uncertainties, principally due to the Doppler effect, and for a comparison with the theory without the uncertainties arising from one-photon QED predictions and nuclear size corrections.

\end{abstract}

\section{Introduction}
Heliumlike heavy ion spectroscopy represents an unique probe of relativistic and Quantum Electrodynamics (QED) effects on the electron-electron interaction in the domain of strong fields.
As compared to a one-electron and many-electron ions, these ions are the simplest multibody systems where theory can make predictions in a rigorous way. 
Here we present the first clear identification of the intra-shell transition $1s2p\, ^3\!P_2 \to 1s2s\, ^3\!S_1$ of He-like uranium performed via X-ray spectroscopy. In addition, using the deceleration capabilities of the ESR storage ring, we performed a differential measurement between this transition and the analog transition $1s^22p\ ^2\!P_{3/2 } \to 1s^22s\, ^2\!S_{1/2}$ of Li-like uranium. 
In this paper we will focus on some aspects of this measurement and in particular on its outlooks.
A more complete description of the experiment can be found in Ref.~\cite{Trassinelli2009b}.

\section{Description of the experimental setup and data acquisition}
The experiment was performed at the GSI experimental storage ring ESR  in August 2007. Here, a H-like uranium beam
with up to $\sim 4 \times 10^7$ ions was stored, cooled, and decelerated to an
energy of 43.57 MeV/u. Excited He-like ions were formed by electron
capture during the interaction of the ion beam with a supersonic
nitrogen gas-jet target. 
At the selected velocity,
electrons are primarily captured into shells with principal
quantum number of $n \leq 20$, which efficiently
populate the $n=2$ $^3\!P_2$ state via cascade feeding. 
This state decays to the $n=2$ $^3\!S_1$ state via an
electric dipole (E1) intra-shell transition (branching ration 30\%)
with the emission of photons of an energy close to 4.51~keV.

\begin{figure}[t]
\begin{minipage}{0.46\textwidth}
\includegraphics[height=6cm]{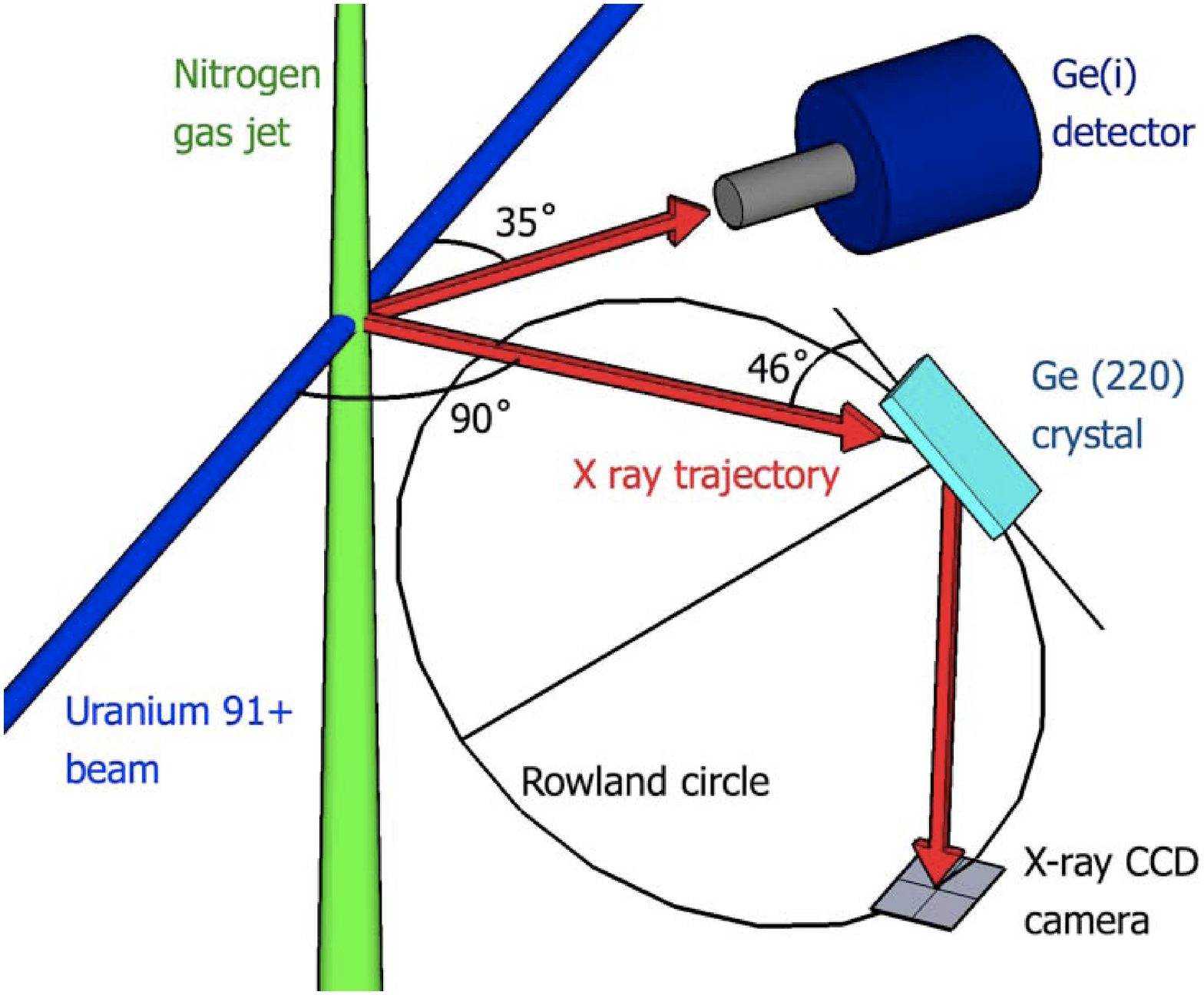}
\caption{\label{setup} Sketch of the experimental arrangement at the gas-jet target of the ESR storage ring. A Ge(i) detector and a focusing Bragg spectrometer are simultaneously viewing the X-ray source defined by the overlap of the circulating ion beam with the gas-jet.}
\end{minipage}
\hfill
\begin{minipage}{0.52\textwidth}
\includegraphics[height=5.7cm]{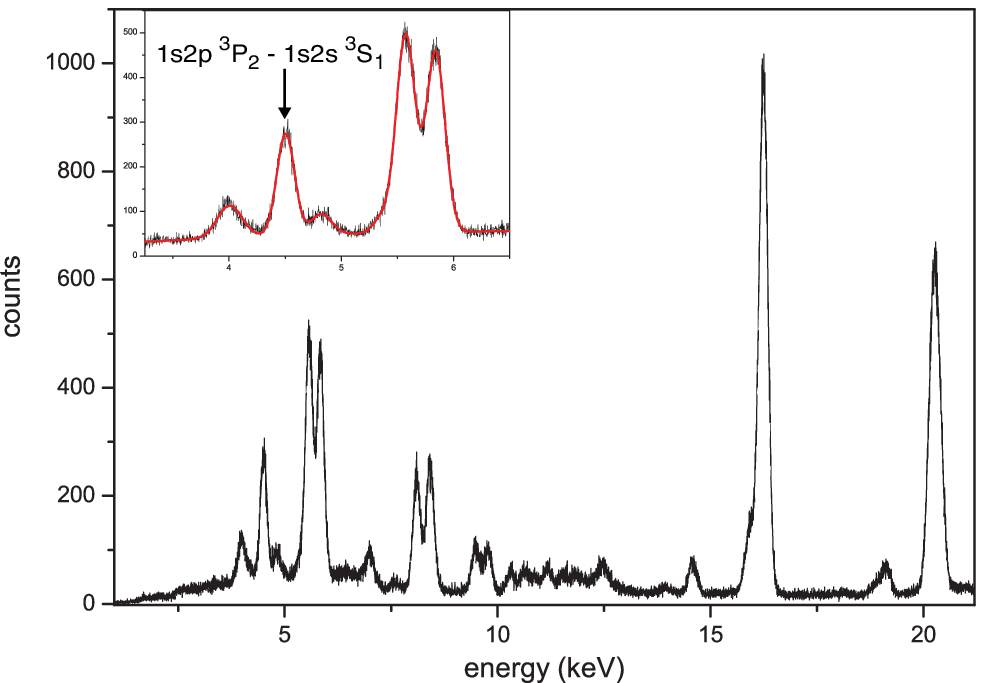}
\caption{\label{Ge-spectrum} The X-ray spectrum, originating from 43.57 MeV/u He-like
uranium ions, as recorded by the Ge(i) detector. The energies correspond to the
emitter frame. The inset shows a magnification of the region around the $^3\!P_2 \to ^3\!S_1$ transition \cite{Trassinelli2009b}.}
\end{minipage}
\end{figure}

For the X-ray detection, we used a standard Ge(i) solid-state detector and a new Bragg spectrometer
specially designed for accurate spectroscopy of fast ions.
The two instruments are complementary: The Ge(i) detector has a high detection efficiency and
covers a wide spectral range with a moderate spectral resolving power. The focusing crystal spectrometer
serves as an accurate wavelength comparator in a narrow wavelength interval.
The Ge(i) solid-state detector and the Bragg crystal 
spectrometer were mounted under observation angles of $35^\circ$ and $90^\circ$, respectively. 
Both instruments were
separated from the ultra-high vacuum of the gas-target chamber by
100$\mu$m-thick beryllium windows transparent for the few keV X rays.
A scheme of the experimental setup is presented in Fig.~\ref{setup}.
The crystal spectrometer \cite{Beyer1991} was mounted in the Johann geometry 
in a fixed angle configuration allowing for the detection of X rays with a Bragg angle $\Theta$ around $46.0^\circ$. 
The spectrometer was equipped with a Ge(220) crystal cylindrically bent, with a radius of curvature of $R=800$~mm, and an X-ray CCD camera (Andor DO420), as position sensitive detector, positioned on the Rowland circle (the focusing position of the Johann spectrometer).
The Rowland-circle plane
of the spectrometer was placed perpendicular to the ion-beam direction. 
In such a configuration the spectral lines appear slanted in the image plane of the spectrometer with their slope proportional to the ion-beam velocity.
For a minimization of the systematic effects due to the ion velocity and alignment uncertainties, the observation angle  $\theta = 90^\circ$ was chosen for the crystal spectrometer.
The value of the ion velocity was selected such that the photon energy, $E$ in the ion frame, was Doppler-shifted to the value $E_\text{lab} = 4.3$~keV in the laboratory frame.
This value of $E_\text{lab}$ was chosen to have the He-like uranium spectral line position on the CCD close to the position of the 8.6 keV $K\alpha_{1,2}$ lines of zinc, which were observed in second order diffraction.
The zinc
lines were used to monitor the spectrometer stability and they were
produced by a commercial X-ray tube and a removable zinc plate
between the target chamber and the crystal. 
Additional information on the experimental setup can be found in Ref.~\cite{Trassinelli2009b}.

For the accurate energy measurement of the He-like intra-shell transition energy, the $1s^22p\, ^2\!P_{3/2} \to  1s^22s\, ^2\!S_{1/2}$
 transition in Li-like U, which has an energy of $4459.37 \pm 0.21$~eV \cite{Beiersdorfer1993,Beiersdorfer1995} was chosen as calibration line. 
Similar to the He-like system, the Li-like ions were obtained by electron capture into He-like
uranium ions. To match the energy of the He-like transition, an
energy of 32.63 MeV/u was selected to Doppler-shift the Li-like transition. 

The data were acquired during a total period of about 4.5 days. 
Survey spectra were recorded by the Ge(i) detector.
Here the $2\,^3\!P_2 \to 2\, ^3\!S_1$ transition was easily identified close to the lines originating from $n' \to n$ transitions (see Fig.~\ref{Ge-spectrum}) with $n =3-4$ \cite{Trassinelli2009b}. 
The measurement with the crystal spectrometer provided a much higher accuracy for the spectral line position. 
In this case, the observable energy range, 
 principally limited by the ion beam diameter and its distance 
from the crystal, was in the order of $4308\pm40$~eV.
For the transition in He-like uranium, a total number of about 300 counts in an effective acquisition time of 
$\sim 24$ hours was accumulated.
For the Li-like ions, about 160 counts in $\sim 5$ hours were recorded.
These spectra are characterized by a very low background 
drastically reduced only
by the energy cuts and cluster analysis of the CCD raw data. 

\begin{figure}[b]
\begin{center}
\includegraphics[width=0.49\textwidth]{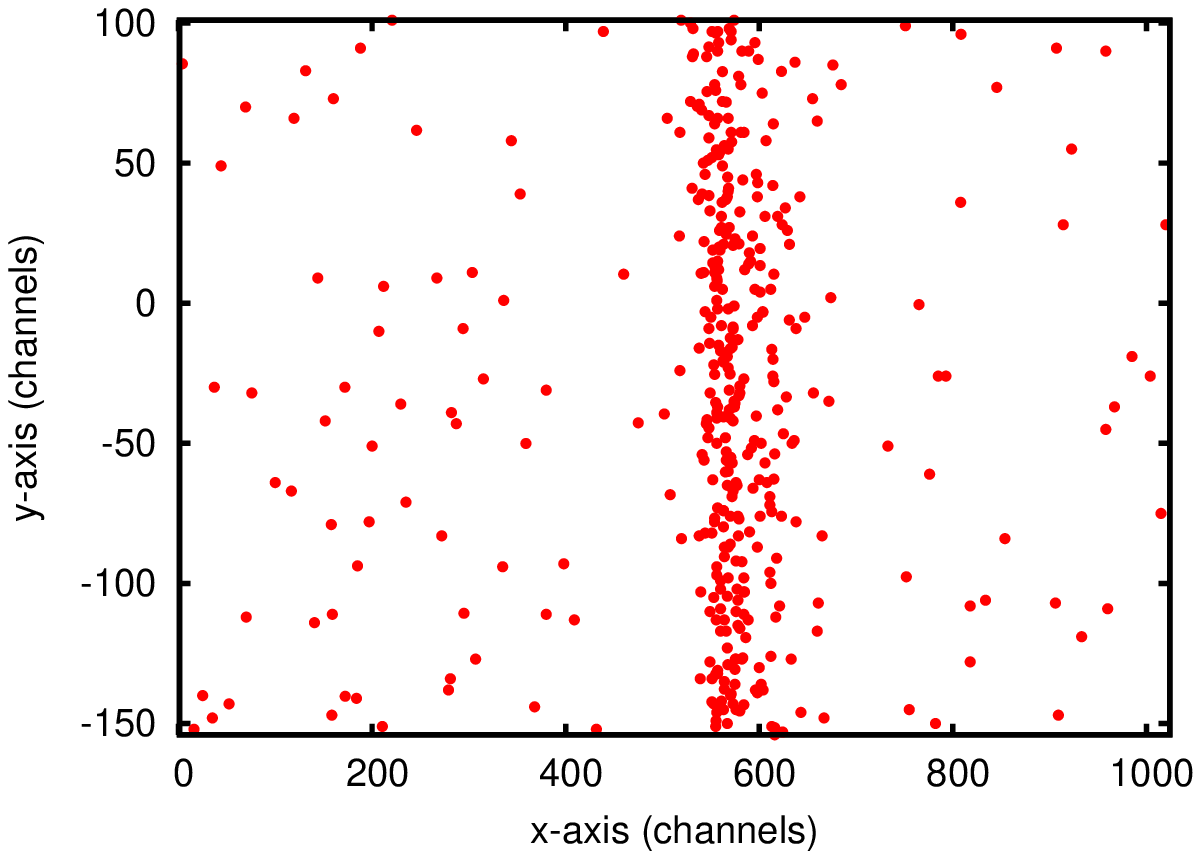}
\includegraphics[width=0.49\textwidth]{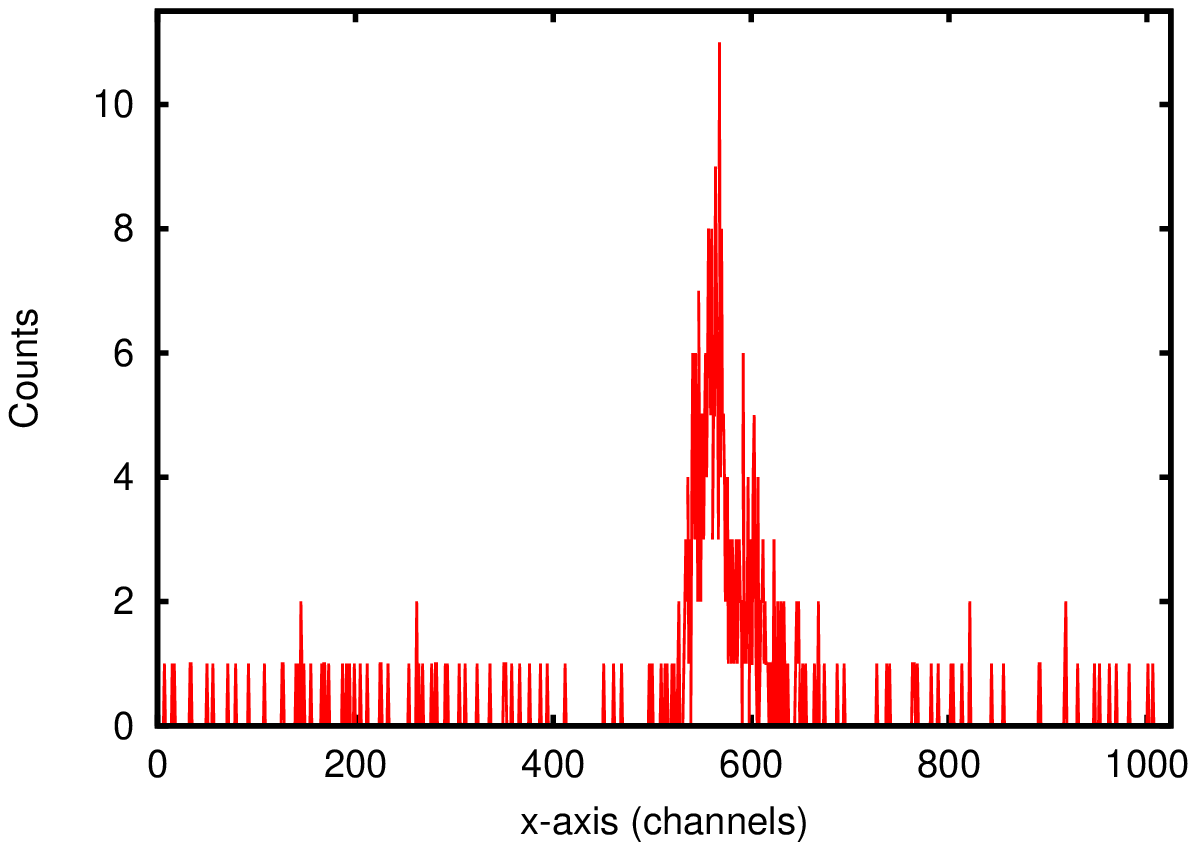}
\end{center}
\caption{\label{spectra} Left: Image of the He-like  uranium intra-shell  transitions  on the Bragg spectrometer position sensitive detector. The transition energy increases with increasing  x-position. 
The slightly negative slope of the line is due to the relativistic velocity of the fast ions. Right: The corresponding high resolution energy spectrum from the projection on the dispersion axis (x-axis) \cite{Trassinelli2009b}. }
\end{figure}

High resolution energy spectra were obtained by projecting the transition lines from the Bragg spectrometer CCD (Fig.~\ref{spectra} left) on the dispersion axis (the x-axis in the figure), after the slope correction due to the Doppler effect (Fig.~\ref{spectra} right).
We note that the shape of the lines corresponding to the fast ion emission is slightly asymmetric. 
Such an asymmetry might be caused by a non-uniform X-ray reflectivity of the crystal surface. After the experiment, a survey of the crystal surface by the X-ray Optics Group from the Institute for Optics and Quantum Electronics in Jena indeed revealed a non-uniformity.

\section{Results and discussions}
Starting from the Bragg law, the energy of the He-like U transition $E_\textrm{He}$ is related to the energy of the calibration line $E_\textrm{Li}$ by the simple formula
\begin{equation}
E_\textrm{He} \approx E_\textrm{Li} 
\frac { \gamma_\textrm{He}} {\gamma_\textrm{Li}} 
\left(  1 +  \frac {\Delta x}{D\ \tan \Theta_\textrm{B}}   \right),
\label{eq:final}
\end{equation}
where $\Delta x = x_\textrm{He} - x_\textrm{Li }$ denotes the
distance between the He- and Li-like U line positions on the CCD and
$D$ the distance between crystal and CCD.
$\gamma_\textrm{He}=1.04677$ and $\gamma_\textrm{Li}=1.03503$ are
the Lorentz factors corresponding to the velocities of the stored H- and
He-like ions, respectively. 
Their values are determined by the
accurately known voltages of the electron cooler of the ESR \cite{Franzke1987}.
From the experimental measurement of $\Delta x$, using Eq.~\eqref{eq:final},
 the energy of the $1s2p\, ^3\!P_2 \to 1s2s\, ^3\!S_1$ transition was determined to be
\begin{equation}
E_\textrm{He} = 4509.71\pm 0.48_\text{stat} \pm 0.86_\text{syst}\; \text{eV},
\label{eq:result}
\end{equation} 
The first uncertainty is statistical and the second one is due to
the systematic uncertainties (quadratically summed).
Principal contributions to the last term are i) the asymmetric response function of the spectrometer, ii) the calibration line and iii) the uncertainty of the observation angle.
All other contributions, due to the accuracy of the electron cooler voltage, the crystal-detector distance, etc., are negligible. A more extended discussion on the systematic uncertainties can be found in Ref. \cite{Trassinelli2009}.
The systematic effect introduced by
this asymmetry has been estimated by comparing the results of the line position
measurement obtained from different approaches (median distribution, and a series of fit adjustments).
This contribution results to be the largest systematic uncertainty of 0.83~eV.
The uncertainty of the observation angle, equal to $0.04^\circ$, contributes with 0.11~eV and it is due to the limited position accuracy of the gas-jet position ($\pm 0.5$~mm) combined with the Doppler shift of the ion emission. We note, the use of the fast Li-like ion transition as calibration, instead of a stationary source, like the Zn K$\alpha$ lines, causes a reduction of such a systematic uncertainty from 0.9 to 0.1~eV \cite{Trassinelli2009}.

\begin{figure}[t]
\begin{center}
\includegraphics[width=0.49\textwidth, trim= 6mm 2mm 6mm 2mm, clip=true]{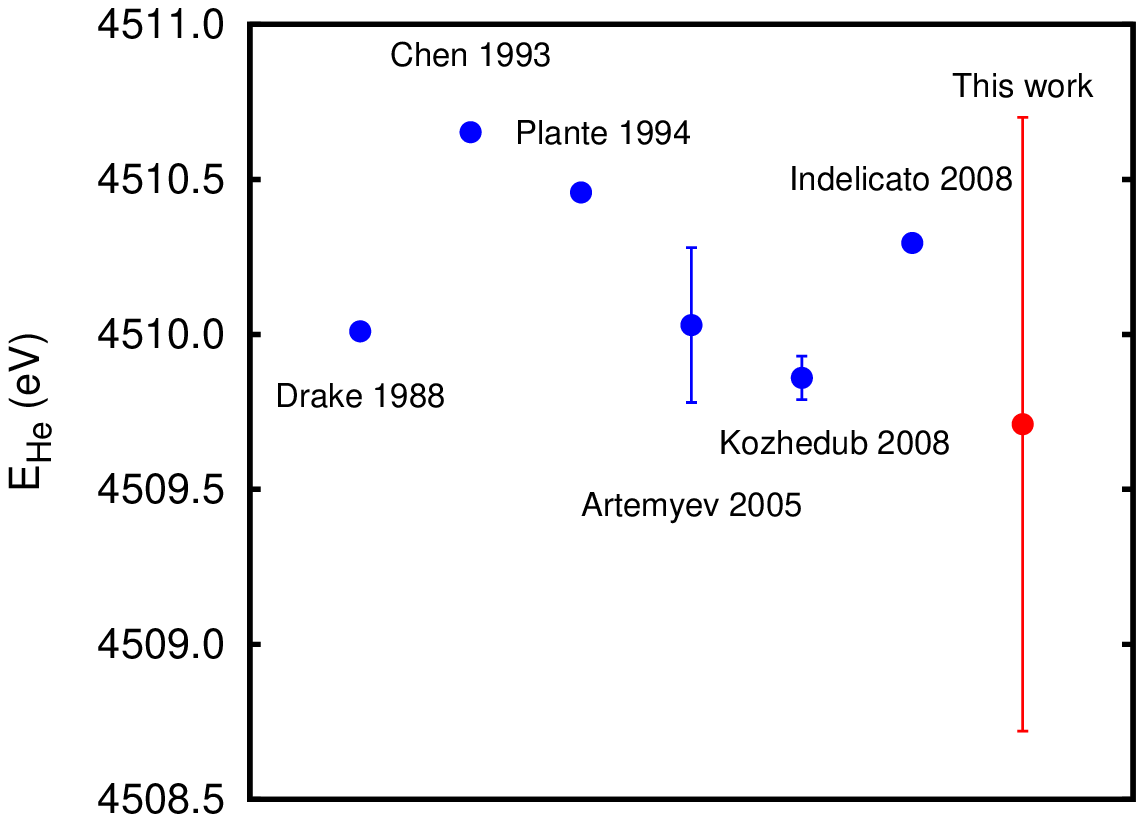}
\hfill
\includegraphics[width=0.49\textwidth, trim= 6mm 2mm 6mm 2mm, clip=true]{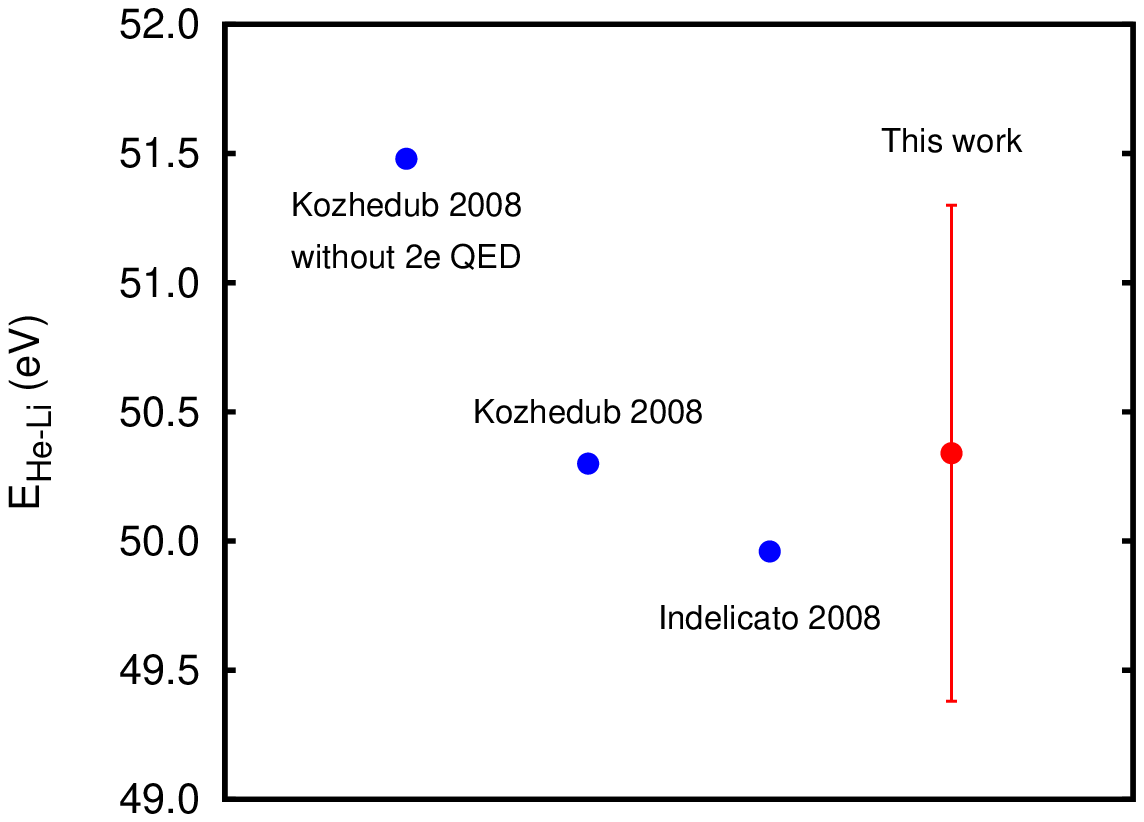}
\end{center}
\caption{\label{exp-th} Comparison between our result and different theory approaches for the He-like U intra-shell transition energy (left) and its relative measurement with respect to the Li-like transition (right). In these figures the reference of the different predictions are:  Drake 1988 \cite{Drake1988}, Chen 1993 \cite{Chen1993}, Plante 1994 \cite{Plante1994}, Artmyev 2005 \cite{Artemyev2005}, Kozhedub 2008 \cite{Kozhedub2008} and Indelicato 2008 \cite{Indelicato2008}.}
\end{figure}

The measured 
transition energy for He-like uranium 
agrees well with  all more recent theoretical predictions,
which reflect different approaches: 4510.30~eV, using a multi-configuration Dirac-Fock calculation with QED corrections included \cite{Trassinelli2009b,Indelicato2008}, and $4509.86 \pm 0.07$~eV, from an \textit{ab initio} calculation \cite{Trassinelli2009b,Kozhedub2008,Artemyev2005} and other older calculations \cite{Drake1988,Chen1993,Plante1994} (see Fig.~\ref{exp-th} (left)).
The accuracy of the present experiment, equal to 0.99~eV, is on the same order as the QED effects on the electron-electron interaction, 0.76~eV for this transition, but is too large to be sensitive to the two-loop QED effects that contributes with  0.20~eV \cite{Kozhedub2008,Artemyev2005}. 

A more significant test of the electron-electron interaction in strong Coulomb fields, comes from the comparison between the measured He- and Li-like transition energy difference and the corresponding theoretical predictions.
From the experimental side, the energy difference eliminates the uncertainty contribution due to the calibration line (0.21~eV), whereas, from the theoretical side, the systematic effects due to the one-electron QED and nuclear size uncertainties are canceled out.
In this case we find
\begin{equation}
E_\textrm{He-Li} = 50.34\pm 0.48_\text{stat} \pm 0.84_\text{syst}\; \text{eV}.
\label{eq:result2}
\end{equation}
In contrast to the absolute value of the transition energy,
the difference carries an experimental systematic uncertainty, peak asymmetry excluded,that is reduced from 0.24 to 0.11~eV, where in practice only the observation angle accuracy contributes. 
This value is in agreement with the theoretical predictions: 49.96~eV \cite{Trassinelli2009b,Indelicato2008} and $50.30 \pm 0.03$~eV  \cite{Trassinelli2009b,Kozhedub2008,Artemyev2005} where the QED effects on the electron-electron interaction contribute with 1.66~eV. For a visual comparison see Fig.~\ref{exp-th} (right), where the role of the two-electron QED effect is presented.
As in the absolute transition energy, the accuracy of the present experiment is on the edge to be sensitive to QED effects.
However, in future experiments, improved and verified analyzer crystals together with an extended acquisition time will allow for more stringent tests of the two-electron QED contributions in heavy highly charged ions. A more extended discussion on the possible outlooks is presented in the next section.

\section{Potential improvements}
From the analysis of the different uncertainty contributions, possible improvements of the present experiment can be clearly inferred. 
As presented in the previous section, the main source of accuracy limitation is due to the quality of the Germanium crystal employed for the Bragg spectrometer. A new crystal carefully selected and X-ray optically characterized could allow to eliminate completely such a systematic effect.
In the relative measurement, the second largest contribution of the systematic uncertainties comes from the alignment accuracy of the crystal spectrometer with respect to the gas-jet position.
This uncertainty can be reduced considerably by the use of two twin Bragg spectrometers at observation angles of $+90^\circ$ and $-90^\circ$. An accuracy better than $0.01^\circ$ could be achieved on the alignment of the two spectrometer with respect to each other.
In this case, the position of the gas-jet with respect to the spectrometers axis can be used for compensation by comparing the energy measurement from each spectrometer. A reduction of the uncertainty from 0.11~eV to about 0.02~eV can be expected.
The use of a second spectrometer will also improve the statistical uncertainty.
A larger crystal radius of curvature will enhance the spectrometer dispersion implying a reduction of instrumental efficiency that can be compensated by a larger position sensitive detector area, limited to $ 26.6 \times 6.6$~mm$^2$ in the present experiment.
Finally, the simplest improvement can come by a longer acquisition time that, we remember, was limited in the present experiment to only 4.5 days. 
A reduction of the statistical uncertainty from 0.5 to 0.1~eV can be expected.

With these upgrades, a future experiment could provide a critical test of the theoretical predictions for the interaction between bound electrons in the presences of a strong Coulomb field where, in particular, the QED contributions could be accurately measured.
In the present experiment we limited to compare He- and Li-like heavy ions. In the future, this study could be extended to additional atomic systems, like Be- and B-like ions, applying the same experimental technique strongly based on the Doppler tuning of the X-ray energy in the laboratory frame.

\section{Conclusions}
In summary, we report the first clear identification 
of the $1s2p\, ^3\!P_2 \to 1s2s\, ^3\!S_1$ transition
in He-like uranium. 
In addition we measured the transition energy 
of such a transition with a
relative uncertainty of $2\times10^{-4}$, 
which is currently the most
accurate test of many-body and QED contributions in excited levels of 
very heavy He-like ions.
Differential measurements between different charge states of the
same fast ion 
pave the way for increased sensitivity via the reduction of the 
systematic uncertainty in both the experimental and the theoretical side.
We also discuss several possible improvements that can be applied to the present experiment to perform more stringent tests on QED and relativistic effects in few-electrons heavy highly charged ions.

\section*{Acknowledgments}
We thank V.~Shabaev, A.N.~Artemyev  and A.~Surzhy\-kov for
interesting discussions and theoretical support. 
We thank O. Wehrhan, H. Marschner and  E. F\"orster for the characterization of the spectrometer crystal.
The close
collaboration and support by the members of the ESR team, the A. von
Humboldt Foundation (M.T.), the DAAD (A.K., No.: A/05/52927) and I3
EURONS (EC contract no. 506065) are gratefully acknowledged.
This work was partially supported by Helmholtz Alliance HA216/EMMI.
Institut des Nanosciences de Paris and Laboratoire Kastler Brossel
are Unit{\'e} Mixte de Recherche du CNRS n$^{\circ}$ 7588 and
n$^{\circ}$ 8552, respectively.

\section*{References}
\providecommand{\newblock}{}

\bibliographystyle{iopart-num}

\end{document}